\documentclass{IOS-Book-Article}     
\usepackage{mathptmx}

\usepackage{amsmath, amsfonts, latexsym}
\usepackage{numprint}
\npdecimalsign{.}
\usepackage{xcolor}
\usepackage{hyperref} 
\usepackage[nameinlink]{cleveref}
\usepackage{threeparttable}
\usepackage{graphicx}
\usepackage{float}
\usepackage{afterpage}
\usepackage{multirow}
\usepackage{needspace}
\usepackage[all]{nowidow}
\usepackage[sort, numbers]{natbib}
\begin{document}
\begin{frontmatter}          
\title{Ontology Reuse: the Real Test of Ontological Design}
\runningtitle{Ontology Reuse: the Real Test of Ontological Design}

\author[A,B]{\fnms{Piotr} \snm{Sowiński}},
\author[B]{\fnms{Katarzyna} \snm{Wasielewska-Michniewska}},
\author[A,B]{\fnms{Maria} \snm{Ganzha}},
\author[B]{\fnms{Marcin} \snm{Paprzycki}},
\author[C]{\fnms{Costin} \snm{Bădică}}
\runningauthor{Sowiński et al.}
\address[A]{Warsaw University of Technology, Warsaw, Poland}
\address[B]{Systems Research Institute, Polish Academy of Sciences, Warsaw, Poland}
\address[C]{University of Craiova, Romania}
\begin{abstract}

Reusing ontologies in practice is still very challenging, especially when multiple ontologies are (jointly) involved. Moreover, despite recent advances, the realization of systematic ontology quality assurance remains a difficult problem. In this work, the quality of thirty biomedical ontologies, and the Computer Science Ontology are investigated, from the perspective of a practical use case. Special scrutiny is given to cross-ontology references, which are vital for combining ontologies. Diverse methods to detect potential issues are proposed, including natural language processing and network analysis. Moreover, several suggestions for improving ontologies and their quality assurance processes are presented. It is argued that while the advancing automatic tools for ontology quality assurance are crucial for ontology improvement, they will not solve the problem entirely. It is ontology reuse that is the ultimate method for continuously verifying and improving ontology quality, as well as for guiding its future development. Specifically, multiple issues can be found and fixed primarily through practical and diverse ontology reuse scenarios.

\end{abstract}


\end{frontmatter}


\section{Introduction}

Ontologies are invaluable in representing domain knowledge; possessing a number of characteristics that make them a tempting choice when designing intelligent information systems~\cite{noy2001}. They are very expressive and flexible, allowing one to describe a wide variety of subjects, in a precise manner. Nowadays, with the introduction of the Linked Open Data principles~\cite{bizer2011linked}, accessing, reusing, and combining ontologies should have become easier than ever. However, in practice, ontology reuse can still be very challenging~\cite{fernandez2019ontologies}. This is especially true when multiple ontologies are involved, where ambiguous or conflicting statements between ontologies often require human intervention in order to be resolved~\cite{cota2020landscape}.

Obviously, high quality of an ontology is vital to ensure its potential reusability across a spectrum of applications. As ontologies model knowledge, any issues in them immediately reflect on the quality of their application. Many criteria for evaluating ontologies were proposed, and there are multiple approaches to their evaluation~\cite{brank2005survey,raad2015survey}. However, individual ontologies use very different evaluation methods, or (in most cases) do not have any evaluation plan in place~\cite{amith2018assessing}. Some approaches are hard to implement in practice -- such as the gold standard method, for which, often, there is no available gold standard to be used. Obviously, expert-based evaluation is resource-intensive and naturally limited in its scope. On the other hand, the automated, more universal methods are usually too general to identify many issues in the ontologies~\cite{al2019towards}.

Overall, as of today, there do not seem to be any established, agreed upon, and widely used standards and methodologies for ontology evaluation. Instead, for years, there have been examples of evaluations that were \emph{tailored to the specific ontology}~\cite{casellas_eval,laadhar2020investigating} or motivated by a \emph{specific, practical use case}~\cite{szmeja2013reengineering,al2019towards,ganzha2017vocabulary}. Lack of a commonly accepted evaluation methodology is one of problems negatively impacting potential reuse and adaptation of ontology-based solutions, as it was investigated, for instance, for the enterprise domain~\cite{wasielewska2021enterprise}.

In earlier work~\cite{sowinski2022topical} an unsupervised, ontology-based method for classifying scientific publications has been described (see Section~\ref{sec:use_case}). During research, when the classifier was to be adapted to different domains, numerous problems with the used ontologies have been noticed. Each of those issues either (i)~increased the amount of work necessary to adapt the classifier to the domain, or (ii)~decreased the accuracy of predictions. These observations motivated further research, which is reported in this contribution.

Specifically, the goal of this work is to demonstrate that the \emph{varied and demanding reusability scenarios} provide the ultimate measure of the quality of an ontological design. Here, the real-world-anchored use case motivates exploring each issue and supports the conclusions drawn from the experiments.

In this context, in the following sections, ontologies from two different fields of science are investigated. To provide additional context to the study, Section~\ref{sec:use_case} briefly describes the publication classification method that serves as the driving use case. Section~\ref{sec:obo_ont} explores thirty diverse biomedical ontologies, related to the field of food safety. Section~\ref{sec:cso_ont} examines a medium-sized Computer Science Ontology.

It should be stressed that all code used in the reported work, as well as additional results and materials, are available on GitHub, under an open-source license~\footnote{\url{https://github.com/Ostrzyciel/ontology-quality-2022}}. Detailed instructions for reproducing the experiments are also provided there.

\section{Use Case Description}
\label{sec:use_case}

The investigation presented in this contribution is driven by a specific use case from our earlier work~\cite{sowinski2022topical} -- an unsupervised method for classifying scientific publications, that aims to be domain-agnostic. Here, all domain knowledge necessary to perform the classification is obtained from ontologies relevant to the pertinent field of science.

The considered method uses the publication's title, abstract, and (optionally) unstructured keywords as an input. It then assigns to the article several entities from the loaded ontologies that best describe its topic. The classification pipeline consists of six steps. 
(1) Using a neural named entity recognition method, possible mentions of entities of interest are identified in the text. 
(2) The ontologies are searched for candidate entities that may match the mention. This is done using a text index. 
(3) The neighborhoods (most closely related entities) of the candidate entities are retrieved from the ontologies. 
(4) Using text embeddings and string matching, the candidate entities are compared to the mention in the text. Here, both the context of the mention (its sentence) and the context of the entity (its neighborhood) are taken into account. This two-way context sensitivity helps the method to correctly classify ambiguous mentions in the text. 
(5) The relations between the identified candidate entities are examined (entity-entity coherence). Then, the least consistent (least strongly connected) candidates are removed. 
(6) For the final result, the highest-scoring and most frequently occurring candidate entities are selected. The result is further enhanced with their parent entities. This allows for classification of publications with topics that are only implied and never mentioned explicitly.

The method relies solely on the ontologies as its source of domain knowledge, utilizing both textual labels (search index, text embeddings) and relations between entities (neighborhood estimation, entity-entity coherence, results enhancement). Unsurprisingly, the method's performance was observed empirically to be largely determined by the quality of the used ontologies. Following the ontology evaluation criteria as described by Raad \& Cruz~\cite{raad2015survey}, accuracy, completeness, conciseness, consistency, and clarity of the ontologies had a direct, significant impact on the classifier's performance. The observed issues are discussed and investigated in detail in the following sections.

\section{Evaluation of OBO Foundry Ontologies}
\label{sec:obo_ont}

Food safety was the first field, to which the publication classification method was applied. As a source of domain knowledge thirty ontologies from the Open Biological and Biomedical Ontology (OBO) Foundry~\cite{smith2007obo} were used, along with SNOMED~CT~\cite{donnelly2006snomed} (full list of used ontologies is available on GitHub\footnote{\url{https://github.com/Ostrzyciel/ontology-quality-2022/blob/main/obo/ONTOLOGIES.md}}). Here, we focus only on the OBO ontologies, which range in size from \numprint{1221} triples for BFO (Basic Formal Ontology) to \numprint{16105832} triples for NCBITaxon. When combined, the thirty ontologies contained \numprint{39156273} triples. These ontologies are highly heterogenous, due to their different goals and being, typically, maintained by independent groups. Most considered ontologies are \emph{developed} in the OBO format\footnote{\url{https://owlcollab.github.io/oboformat/doc/GO.format.obo-1\_4.html}}, which is then \emph{translated} into the standard Web Ontology Language (OWL), to facilitate reuse.

The issue of reliably maintaining quality over independently developed ontologies is an active area of research, with tools such as ROBOT~\cite{jackson2019robot} or the OBO Dashboard~\cite{Jackson2021OBOFI}, attempting to tackle it. OBO Dashboard, in particular, performs a wide selection of tests on ontologies, essentially operationalizing the review activities that would otherwise be performed by a human. However, the set of tests is limited -- for example the validity of cross-ontology references is not checked. Most of the issues have to be fixed manually, which proves difficult in practice, especially for the rarely-updated ontologies. Nevertheless, it can be stipulated that OBO Foundry is making significant efforts to assure the high quality of its ontologies. Keeping this in mind, let us report issues that have been found.

\subsection{Ease of Access}

Before the ontologies could be used, they had to be first downloaded and loaded into a triple store. All ontology files downloaded from the OBO Foundry had the \texttt{.owl} extension, but their actual format varied. Most were distributed in the RDF/XML format, which is in line with the official OBO Foundry guidelines. However, the Common Anatomy Reference Ontology (CARO) used the OWL functional syntax instead, and had to be converted into RDF before being loaded. The Food-Biomarker Ontology (FOBI) was distributed in the OWL/XML format, which could not be converted to RDF/XML, due to invalid element errors. Thus, for FOBI, an earlier release was used that did not exhibit this issue and could be successfully converted to RDF.

It is worth pointing out that although these issues may appear as \textit{minor annoyances}, they have significant implications. Inconsistency in file formats makes automatic reuse of these ontologies much harder or even impossible, due to the need for manual conversion of the files.

\subsection{Rarely Used Properties}

The publication classifier requires the user to assign weights to all object properties that may be useful for discovering how different entities are interrelated (steps 3 and 5 of the method). Similarly, weights are assigned to text properties, which describe the entities (step 2). To ascertain these weights, an obvious first step is to retrieve a list of all properties, sorted by the number of times they were used. Although OBO Foundry ontologies strive for reuse and have common upper ontologies defining relations, in the resulting lists many rarely used properties were found, often seeming like mistakes. Thus, those properties that had at most ten unique uses across all ontologies have been manually inspected.

It was found that out of 140 rarely-used properties, 74 were in some way erroneous, while being used across 278 unique triples. The most common type of error (51 properties, 212 occurrences) were undefined properties in the oboInOwl namespace~\cite{moreira2009ncbo}. These are, most likely, errors in translation from the OBO format to OWL. Other issues included mistaken ontology prefixes, typos, and invalid URIs. These issues were found in 24 out of thirty ontologies.

\subsection{Property Value Type Mismatch}

Next, the classifier requires dividing properties into URI-valued and literal-valued. This turned out to be a non-trivial task, as many properties were used inconsistently -- sometimes their objects were URIs, and sometimes they were literals. 

A list of all properties was created, with the number of times each was used in triples in which the object is an URI, a blank node, or a literal. For manual review, properties that exhibited conflicting usage patterns were selected. Most did not have an \texttt{rdfs:range} property, which would define formally their set of possible values. In fact, some properties were not defined at all, in any ontology. In other cases, the range was specified very broadly, allowing both URIs and literals. This presented a major challenge in distinguishing between erroneous and valid annotations. In such cases it had to be decided ``intuitively'' whether a given use is valid or not. The decision was made on the basis of scarce documentation, or other occurrences of the same property. After establishing the valid usage patterns, lists of erroneous occurrences for each property were obtained, using SPARQL queries.

In the results, the \texttt{oboInOwl:hasDbXref} property was excluded, as it is discussed separately in the following subsection. Overall, \numprint{12296} erroneous triples were identified in 20 ontologies. Often, in places where an URI-typed value was expected, a string literal representing that URI was present. Thus, the value itself was valid, but the datatype did not match.  In other cases, various ontologies had different conventions for applying the same property. For some metadata properties, e.g., those for marking authorship, significant inconsistencies in their use were observed. However, it was decided not to mark them as errors, due to our lack of knowledge as to which convention should be applied, in a given case. Nonetheless, the lack of a common convention for representing metadata makes understanding and reusing ontologies considerably harder.

\subsection{Inconsistent Cross-ontology References}

Food safety integrates a wide variety of topics, from microbiology, to chemistry, and human diseases. Therefore, it is crucial to ensure that the imported ontologies are well-connected. Such connections, in OBO Foundry, can be either partial imports, (where an entity from one ontology is imported into another), or cross-ontology references, made with various types of properties. When examining inter-ontology references, it was observed that their usage patterns varied and their targets were hard to resolve automatically. Thus, all occurrences of cross-ontology references were retrieved and their targets analyzed. For these references, the dominant property is \texttt{oboInOwl:hasDbXref} (\numprint{3809415} uses), but some SKOS~\cite{Miles:09:SSK} properties are also used. There were 259 occurrences of \texttt{skos:*Match} properties, of which only three had valid URIs as values. Others had literals of external identifiers, e.g., \texttt{MESH:C536189}.

The SKOS references were not included in further analysis. Instead, two standard OBO properties from the \texttt{oboInOwl} namespace were analyzed: \texttt{hasDbXref} and \texttt{hasAlternativeId}. Their values are supposed to be strings, having the form of \texttt{Namespace:Identifier}, with the namespace (usually) corresponding to an OBO Foundry ontology. All triples (\numprint{3908752}) containing these properties were retrieved. Then, the obtained reference values were processed and classified using a custom algorithm. The found namespace identifiers were cross-referenced with the publicly available list of all OBO Foundry ontologies\footnote{\url{https://obofoundry.org/registry/ontologies.yml}}, to filter out valid references. Out of all references, \numprint{52122} (1.3\%) were URIs, not identifiers. Among them only 112 were valid URIs of entities in OBO Foundry. Others pointed to very diverse resources, e.g., Wikipedia, GitHub, online databases, scientific publications, structured vocabularies, and various other websites.

Among non-URI references, \numprint{187167} (4.8\%) referred to OBO Foundry ontologies. Others, usually, pointed to external biomedical databases and structured vocabularies, such as MeSH or SNOMED~CT. However, multiple prefixes were used for some databases, (e.g., \texttt{SCTID}, \texttt{SMID}, \texttt{SNOMEDCT\_US}, \texttt{SNOMEDCT\_US\_2021\_09\_1} for SNOMED~CT), which complicates translating the identifiers into URIs. Using the recently created Bioregistry~\cite{charles_tapley_hoyt_2020_4404608}, it was possible to resolve prefixes for another \numprint{3013315} references (77.1\%). Nonetheless, the identified databases form an eclectic collection, including the Unified Medical Language System, PubMed, English Wikipedia, ISBN, ORCID, and Google Patents. This shows that the cross-reference properties are used for a wide variety of purposes. Obviously, this complicates their potential reuse. Moreover, the references did not always conform to the \texttt{Namespace:\allowbreak Identifier} format, with some using the underscore (\texttt{\_}) as a separator instead of a colon. For OBO identifiers, there were 19 such cases, while for Bioregistry-resolved prefixes \numprint{26572} were identified. Finally, 214 ($<0.01$\%) references pointed to empty RDF blank nodes. These are, most likely, the results of errors during modeling, or translation into OWL. All these findings are summarized in~Table~\ref{tab:obo_xref}.
\begin{table}[htb]
    \centering
    \caption{Cross-ontology references in OBO Foundry\label{tab:obo_xref}}
    \begin{tabular}{l l r r}
        \hline
        \multicolumn{2}{c}{\textbf{Type}} & \textbf{\# references} & \textbf{\% total} \\
        \hline
        \multicolumn{2}{l}{\textbf{URI}} & \numprint{52122} & 1.33\% \\
        & valid OBO URI & 112 & ~$<0.01$\% \\
        & en.wikipedia.org & \numprint{17365} & 0.44\% \\
        & orcid.org & \numprint{8085} & 0.21\% \\
        & langual.org & \numprint{5592} & 0.14\% \\
        & other & \numprint{20968} & 0.54\% \\
        \hline
        \multicolumn{2}{l}{\textbf{Textual}} & \numprint{3856416} & 98.66\% \\
        & recognized OBO identifier & \numprint{187167} & 4.79\% \\
        & recognized other Bioregistry prefix & \numprint{3013315} & 77.09\% \\
        & unknown & \numprint{655934} & 16.78\% \\
        \hline
        \multicolumn{2}{l}{\textbf{Empty blank node}} & \numprint{214} & ~$<0.01$\% \\
        \hline
        \multicolumn{2}{l}{\textbf{Total}} & \numprint{3908752} & 100.00\% \\
        \hline
    \end{tabular}
\end{table}

It is worth mentioning that the OBO ontology format has a number of header tags (\texttt{treat-xrefs-as-*}) that allow one to more precisely describe how to treat cross-references, on a per-ontology basis. This information, however, is lost in the translation to OWL, and thus largely inaccessible.

The (in)consistency of OBO cross-ontology references has also been explored by Laadhar et al.~\cite{laadhar2020investigating}, who noted the overwhelming variety of uses for the cross-reference properties, which hampers reuse. Their work gives valuable suggestions of possible improvements, which augment the discussion presented here.
\begin{table}[htbp]
    \centering
    \begin{threeparttable}
    \caption{Summary of issues found in OBO Foundry ontologies\label{tab:obo_summary}}
    \begin{tabular}{l r r r r r}
        \hline
        \textbf{Ontology} & \textbf{Rare prop.}\tnote{1}~ & \textbf{Prop. obj.}\tnote{2}~ & \textbf{Xref: blank}\tnote{3}~ & \textbf{Xref: URI}\tnote{4}~ & \textbf{Xref: unk.}\tnote{5}~ \\
        \hline
        AEO & \numprint{4} & \numprint{0} & \numprint{0} & \numprint{10} & \numprint{136} \\
        AGRO & \numprint{18} & \numprint{51} & \numprint{0} & \numprint{1266} & \numprint{6710} \\
        APOLLO-SV & \numprint{4} & \numprint{308} & \numprint{214} & \numprint{2} & \numprint{21} \\
        BFO & \numprint{0} & \numprint{0} & \numprint{0} & \numprint{0} & \numprint{0} \\
        BTO & \numprint{3} & \numprint{0} & \numprint{0} & \numprint{0} & \numprint{3479} \\
        CARO & \numprint{1} & \numprint{6} & \numprint{0} & \numprint{380} & \numprint{1800} \\
        CHEBI & \numprint{12} & \numprint{0} & \numprint{0} & \numprint{0} & \numprint{313736} \\
        CL & \numprint{38} & \numprint{236} & \numprint{0} & \numprint{2297} & \numprint{34296} \\
        DOID & \numprint{2} & \numprint{2} & \numprint{0} & \numprint{1} & \numprint{12824} \\
        DRON & \numprint{9} & \numprint{6} & \numprint{0} & \numprint{0} & \numprint{35148} \\
        EHDAA2 & \numprint{3} & \numprint{0} & \numprint{0} & \numprint{5} & \numprint{67} \\
        ENVO & \numprint{3} & \numprint{1612} & \numprint{0} & \numprint{3299} & \numprint{1649} \\
        FOBI & \numprint{5} & \numprint{0} & \numprint{0} & \numprint{0} & \numprint{0} \\
        FoodOn & \numprint{0} & \numprint{5702} & \numprint{0} & \numprint{8416} & \numprint{6329} \\
        GAZ & \numprint{0} & \numprint{6} & \numprint{0} & \numprint{0} & \numprint{25505} \\
        GO & \numprint{1} & \numprint{2536} & \numprint{0} & \numprint{354} & \numprint{118473} \\
        HP & \numprint{45} & \numprint{313} & \numprint{0} & \numprint{3520} & \numprint{28386} \\
        IAO & \numprint{0} & \numprint{22} & \numprint{0} & \numprint{0} & \numprint{0} \\
        MP & \numprint{47} & \numprint{388} & \numprint{0} & \numprint{15253} & \numprint{37229} \\
        NCBITaxon & \numprint{0} & \numprint{0} & \numprint{0} & \numprint{0} & \numprint{0} \\
        OBI & \numprint{0} & \numprint{1295} & \numprint{0} & \numprint{0} & \numprint{0} \\
        PATO & \numprint{13} & \numprint{96} & \numprint{0} & \numprint{3485} & \numprint{17144} \\
        PCO & \numprint{3} & \numprint{19} & \numprint{0} & \numprint{9} & \numprint{41} \\
        PECO & \numprint{2} & \numprint{0} & \numprint{0} & \numprint{0} & \numprint{685} \\
        PO & \numprint{3} & \numprint{24} & \numprint{0} & \numprint{3} & \numprint{6547} \\
        RO & \numprint{2} & \numprint{35} & \numprint{0} & \numprint{0} & \numprint{15} \\
        SYMP & \numprint{2} & \numprint{0} & \numprint{0} & \numprint{1} & \numprint{449} \\
        Uberon & \numprint{87} & \numprint{375} & \numprint{0} & \numprint{23845} & \numprint{14627} \\
        UO & \numprint{9} & \numprint{0} & \numprint{0} & \numprint{0} & \numprint{0} \\
        XCO & \numprint{3} & \numprint{0} & \numprint{0} & \numprint{0} & \numprint{494} \\
        \hline
        \textbf{All} & \textbf{\numprint{278}} & \textbf{\numprint{12296}} & \textbf{\numprint{214}} & \textbf{\numprint{52122}} & \textbf{\numprint{655934}} \\
        \hline
    \end{tabular}
    \begin{tablenotes}
        \item[1] Invalid occurrences of rarely-used properties.
        \item[2] Property object type mismatch (URI instead of literal or vice versa).
        \item[3] Cross-references pointing to blank nodes.
        \item[4] Cross-references pointing to URIs instead of identifiers.
        \item[5] Non-resolvable cross-reference identifiers.
    \end{tablenotes}
    \end{threeparttable}
\end{table}

\subsection{Summary}
\label{subsec:obo_summary}

In summary, \numprint{720844} issues were found across thirty OBO Foundry ontologies (treated jointly). These issues, divided into five categories, and broken down by the source ontology, have been presented in~Table~\ref{tab:obo_summary}. Note that due to the use of partial imports (cf. MIREOT~\cite{courtot2011mireot}), some ontologies overlap. Thus, the actual number of issues in a given category may be lower than the sum of issues encountered in individual ontologies. It is also worth noting that \emph{only two ontologies} were found to be entirely problem-free, from the perspective of the considered use case. These were BFO and NCBITaxon.

\section{Evaluation of the Computer Science Ontology}
\label{sec:cso_ont}

The same classifier was adapted to the Computer Science Ontology (CSO)~\cite{salatino2018computer}, which models research topics in computer science. Its main areas of application include publication classification and research trend detection. It uses a small set of semantic properties, extending the SKOS data model~\cite{Miles:09:SSK}. 
It also contains references to entities in external knowledge bases (KBs) such as DBpedia~\cite{auer2007dbpedia}, Wikidata~\cite{vrandevcic2014wikidata}, and YAGO~\cite{tanon2020yago}, using the \texttt{owl:sameAs} property. The most recent version of CSO (3.3) contains \numprint{14290} topics and \numprint{163470} triples, and was used in all experiments.

The ontology was generated automatically, using the Klink-2 algorithm~\cite{osborne2015klink}. Relationships in some research areas (Semantic Web and Software Architecture) were manually reviewed by experts, but most of the burden of ontology quality control and maintenance was left to the community. To facilitate this, the authors of CSO built the \textit{CSO Portal}\footnote{\url{https://cso.kmi.open.ac.uk/}}, which collects suggestions for improving the ontology.

In the following subsections, we examine various issues found in CSO. 
To assure that the results properly identify the magnitude of individual issues, each experiment was performed independently, starting from a fresh copy of the ontology. 

\subsection{Synonym Description Structure}
\label{subsec:cso_structure}

In CSO, alternative names of one concept are treated in such a way that each synonym is its own entity of type \texttt{cso:Topic}. Clusters of equivalent topics are connected using the \texttt{cso:related\allowbreak Equivalent} and \texttt{cso:preferential\allowbreak Equivalent} properties. There are \numprint{11187} such clusters. This structure is in a stark contrast to most other ontologies, which usually attach alternative names to a single entity, using text properties (cf. Gene Ontology and other OBO ontologies~\cite{ashburner2000gene}, Open Research Knowledge Graph~\cite{jaradeh2019open}, YAGO~\cite{tanon2020yago}).

\begin{figure}[htb]
\includegraphics[width=10cm]{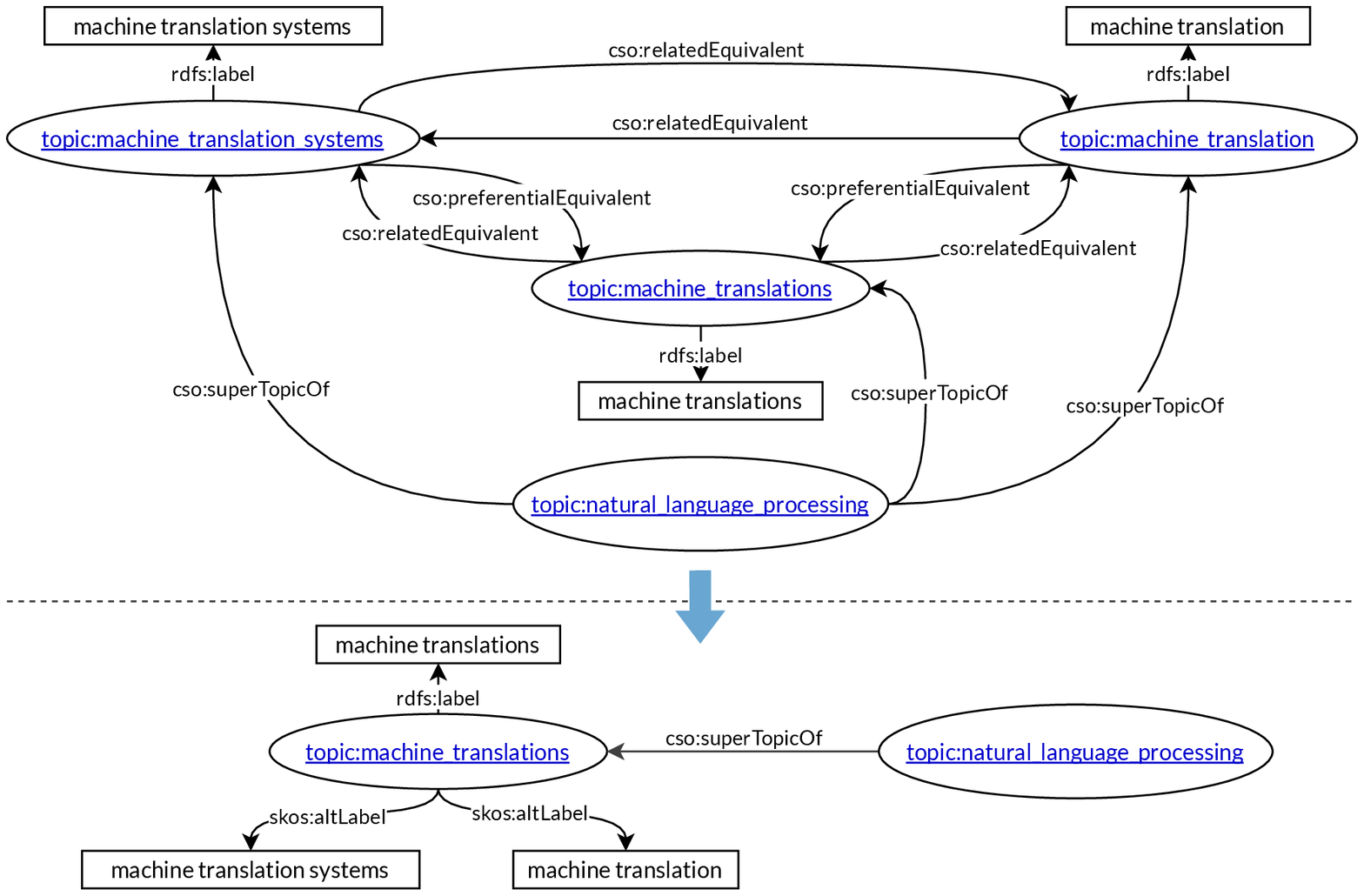}
\centering
\caption{An example of the changes made in CSO}
\label{fig:cso_synonyms}
\end{figure}

When processing the ontology, it was found to be laborsome to work with this structure. For example, when displaying the results of a query searching for topics, one would have to filter out those that are not marked as \emph{preferential}. Similarly, adapting the publication classification method to CSO would be impossible without writing additional logic to handle the atypical structure. Therefore, CSO was transformed using SPARQL 1.1 updates into an ontology with synonymous topics merged into a single entity. Here, synonyms were attached using the \texttt{skos:altLabel} property. The transformation changed the structure of the ontology, but it did not affect its information content. The original \numprint{163470} triples were transformed into \numprint{88227} triples, a 46\% decrease. An example of the applied transformation can be found in Figure~\ref{fig:cso_synonyms}.

\subsection{Externally Inconsistent References}
Next, CSO's references to external KBs were examined. The goal was to be able to use federation to provide additional information to the classifier. However, multiple references were found to point to entities very distant from computer science, such as musicians or rivers. Assuming that in the external KB there should not be many connections between, e.g., popular music and computer science, one could attempt to find the ``outlier'' entities that are not well connected with the rest. These entities will, most likely, be invalid references.

A list of all DBpedia entities referenced from CSO was obtained. Then, for each entity, a list of links to other DBpedia entities was retrieved. This data was used to construct an undirected wikilink graph that was analyzed using the NetworkX library~\cite{hagberg2008exploring}. Three approaches to identifying outliers were used in a sequence:
\begin{itemize}
    \item[\textbf{T1}] Find all connected components (CCs) in the graph. Add the vertices from all except the largest CC to the set of candidate outliers (result set). Afterwards, remove these vertices from the graph, to leave only one CC.
    \item[\textbf{T2}] Identify and remove bridges in the graph, increasing the number of CCs~\cite{BollobasBela2013MGT}. Add the vertices from the newly created CCs to the result set, and then remove them from the graph.
    \item[\textbf{T3}] Find ``communities'' in the graph, using the label propagation algorithm~\cite{cordasco2010community}. Add the vertices from these communities to the result set.
\end{itemize}

The outlier detection method identified 140 potentially erroneous alignments, which were then manually evaluated by a reviewer, using a custom web application. As the errors were usually easily discernible and there was little room for ambiguity, only one reviewer was included in this experiment. Here, each alignment has been marked as \emph{valid}, \emph{invalid} or having a different issue, such as being out of scope of computer science entirely. Table~\ref{tab:ex1} presents the results of the review, broken down by the tactic that led to marking the alignment as ``suspect''. It can be observed that tactics \textbf{T1} and \textbf{T2} were highly effective in discovering potentially erroneous alignments. The false positive rate was relatively low, at only 25\% of all suspected entities not having any issues. Most occurrences in the \emph{other~issues} category referred to out-of-scope entities, which explains why their external references were inconsistent with others.
\begin{table}[htb]
    \centering
    \caption{DBpedia alignments manual review results\label{tab:ex1}}
    \begin{tabular}{c r r r r}
        \hline
        \textbf{Verdict} & \textbf{\# total} & ~\textbf{\# T1} & ~\textbf{\# T2} & ~\textbf{\# T3} \\
        \hline
        invalid & 74 & 47 & 27 & 0 \\
        other issues & 31 & 14 & 15 & 2 \\
        valid & 35 & 15 & 13 & 7 \\
        \hline
        \textbf{Total} & 140 & 76 & 55 & 9 \\
        \hline
    \end{tabular}
\end{table}

\subsection{Missing References to the Corresponding Knowledge Bases}
CSO contains references to several external KBs, most of which have corresponding entities for each other. For example: each DBpedia entity refers to an English Wikipedia article, and each such article has a corresponding entity in Wikidata. Thus, one may expect there to be a 1:1 mapping between CSO's references to DBpedia and Wikidata. To examine this, a series of SPARQL queries was performed, looking for instances where this 1:1 mapping was not the case.

The method did not identify any missing DBpedia references. However, it did find 31 topics with missing references to 13 unique Wikidata items. Fixing the issue is trivial, and a patch in the form of a Turtle file\footnote{\url{https://github.com/Ostrzyciel/ontology-quality-2022/blob/main/cso/3_missing_refs/results/suggestion.ttl}} was prepared.

\subsection{Logically Invalid References}
\begin{figure}[htb]
\includegraphics[width=10cm]{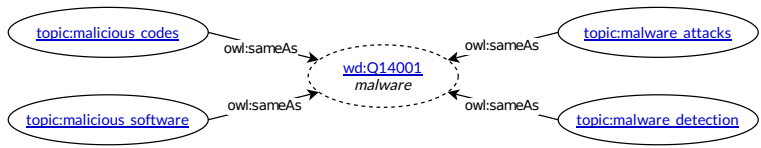}
\centering
\caption{References to external KBs in CSO}
\label{fig:cso_external}
\end{figure}

The publication classifier uses relationships contained in the ontology to determine the degree of similarity between entities. Each property type is assigned a weight, which should be low for very close conceptual connections (e.g., identity), and high for the more ``remote'' relationships. When assigning the weights, one can assume that the \texttt{owl:sameAs} relation, according to its definition, should represent identity (and thus, have zero weight). However, CSO uses the \texttt{owl:sameAs} property for referencing external knowledge bases. This property is defined as reflexive, symmetrical, and transitive~\cite{bechhofer2004owl}, which implies that if two different concepts were to be aligned to the same external entity, they could be deduced to be identical, which is a logic violation. Multiple examples of such problems have been noticed (see, Figure~\ref{fig:cso_external}). 
Here, for example, \emph{malicious software} and \emph{malware detection} can be reasoned to be identical, even though in the ontology they are (as expected) completely separate concepts.

To find all such occurrences, the set of \texttt{owl:sameAs} relations in CSO was expanded, using a reasoner that exploits the reflexive, symmetrical, and transitive characteristics of this property. A SPARQL query was used to obtain a list of CSO topic pairs, marked as identical (\texttt{owl:sameAs}), but not belonging to the same cluster of synonyms. A total of \numprint{7457} unique URI pairs was found to be incorrectly marked as identical. This issue concerned \numprint{3250} unique topic URIs, out of \numprint{14290} CSO topics in total.

For comparison, in the BioPortal, which hosts multiple biomedical ontologies, the alignments that were constructed with the LOOM lexical matching algorithm~\cite{ghazvinian2009creating} are marked using the \texttt{skos:closeMatch} property, chosen due to its lack of transitiveness, which helps avoid similar issues.

\subsection{Intra-cluster Alignment Inconsistencies}
CSO's approach to synonym clusters requires caution to avoid inconsistencies in references to external KBs. For example, having topics $A$ and $B$ from a single synonym cluster, one would expect that if $A$ is aligned to an external entity $X$, then also $B$ is aligned to $X$. Moreover, CSO topics from one cluster should not be aligned to multiple different external entities from a single external KB.

To validate these assumptions, SPARQL queries were performed to find DBpedia alignments for each synonym cluster. The information was then processed with a script to find clusters that violate either assumption. The method identified 130 synonym clusters that contained more than one unique DBpedia reference. Of those, 124 clusters had two unique references, and six clusters had three. Moreover, there were 962 entities with a DBpedia reference missing, spread over 752 synonym clusters. Adding these missing references is easy and an appropriate ontology patch was prepared\footnote{\url{https://github.com/Ostrzyciel/ontology-quality-2022/blob/main/cso/5_intra_ref_consistency/results/suggestion.ttl}}. In total, 882 out of \numprint{2225} (39.6\%) larger-than-one synonym clusters exhibited some external reference consistency issues.

\subsection{Term Conflation}
When debugging the outputs of the publication classifier, several synonym clusters were noticed to include conflated and erroneous terms. Finding more such issues systematically would require the intervention of human experts. To narrow down the set of ``suspected'' clusters, natural language processing techniques can be used to determine the degree of similarity between terms. Then, the least coherent synonyms could be inspected manually by human reviewers.

Hence, using SPARQL queries, a list of all synonym clusters was obtained, along with the corresponding topic labels. These labels were then encoded into vectors using a sentence embedding model \texttt{all-mpnet-base-v2}\footnote{\url{https://huggingface.co/sentence-transformers/all-mpnet-base-v2}}, based on the MPNet transformer~\cite{song2020mpnet}. Next, in each cluster, an all-pairs similarity matrix was computed between the encoded labels (using cosine similarity, however, any other similarity measure could have been used). For each topic, the mean and the standard deviation of its similarity to other topics were computed. Finally, clusters that contained at least three topics and had low mean and standard deviation of similarity to other topics were selected for review. This approach was designed to find clusters with topics that are \emph{systematically} inconsistent, and thus most likely to be wrong.

For manual review, suspected clusters were evaluated by three experts. Available ratings were: \emph{definitely good}, \emph{probably good}, \emph{not sure}, \emph{probably wrong}, and \emph{definitely wrong}. The order of clusters was picked at random, independently for each expert. The experts were allowed to use external sources to aid their work, and were instructed to simply use their best judgment when making the decision.

The intra-cluster similarity method identified 115 clusters of potentially invalid synonyms. Using the simple majority vote rule, 84 of these clusters were deemed \emph{wrong} by the reviewers. Moreover, 95 clusters were marked as such by at least one reviewer, and all three reviewers agreed that 58 clusters are wrong. To investigate inter-reviewer agreement, scores were converted to an ordinal scale from -2 to 2, with \emph{definitely wrong} corresponding to -2, and \emph{definitely good} corresponding to 2. Mean scores and standard deviations for each reviewer are presented in~Table~\ref{tab:ex5}. It can be observed that reviewers had significantly different score distributions (Fleiss' $\kappa = 0.501$, Krippendorff's $\alpha = 0.474$). This indicates moderate agreement, which is desirable when seeking varied opinions on the synonymy of terms.
\begin{table}[hbt]
    \centering
    \caption{Term conflation: reviewer's scores distributions\label{tab:ex5}}
    \begin{tabular}{c r r}
        \hline
        \textbf{Reviewer} & \textbf{Mean score} & \textbf{St. dev.} \\
        \hline
        Reviewer 1 & -1.16 & 1.60 \\
        Reviewer 2 & -0.85 & 1.67 \\
        Reviewer 3 & -0.13 & 1.79 \\
        \hline
    \end{tabular}
\end{table}

It was found that, in CSO, frequently conflated were terms surrounding a single subject. Here, as an example one can consider \emph{classification system}, \emph{classification tasks}, \emph{classification results}. Other synonyms seemed of little use, introducing only singular/plural variations, or different variants of hyphen placement. Abbreviations were also occasionally mixed into the labels -- for example \emph{neural network (nn)} exists as a single term.

Additionally, the reviewers made plentiful remarks about scrutinized clusters. In particular, many out-of-scope clusters were found, frequently from the fields of pedagogy and genetics. This is consistent with the observations of Han et al.~\cite{han2020wikicssh}, who point out that CSO fails to distinguish computer science topics from those of related fields.

\subsection{Summary}

Overall, in above experiments, \numprint{3137} synonym clusters and \numprint{4287} topics were found to be affected by at least one issue. For cases where the fix is unambiguous, suggested patches in the form of Turtle files were prepared and made available.

\section{Discussion and Concluding Remarks}
\label{sec:conclusions}

Over the course of a use case-driven investigation, 31 heterogeneous ontologies were scrutinized with regard to their quality. Not only use case-specific issues were found, but also those related to basic principles of ontology design, clarity and consistency.

Overall, OBO Foundry ontologies present an extremely varied landscape of sizes, quality assurance methods, and stakeholders involved. Thus, it is interesting to observe that out of 30 studied ontologies, 28 were found to involve at least one class of problems. Moreover, in majority of cases, found problems concerned a very large number of triples (these were not ``localized problems'' but clear cases of systemic ones). One method to tackle them would be, for OBO Foundry, to further expand upon their set of guidelines, for example by specifying, which properties to use for describing metadata. Defining formally the ranges of properties could also prove valuable, as it would enable automatic tools to check for potential problems. Cross-ontology references also require improvements, to better conform with Linked Data standards and have clearer meaning. Finally, the OBO to OWL translation should be further improved, to better preserve semantic information. 

In the case of CSO, it was found that some issues stem from the decisions made when designing its overall structure, while others can be attributed to the non-exhaustive quality assurance of the automatically generated ontology. It can be hoped that the reported results and suggested fixes will be valuable to the further development of CSO.

In the experiments, the usefulness of natural language processing methods and network analysis for ontology quality assurance were clearly shown. Most importantly, these techniques can greatly reduce the amount of human labor required to perform the evaluation. Thus, ontological engineers should not be afraid to broaden their toolboxes with less conventional techniques. This is particularly the case since the \textit{joint} use of approaches described above can substantially improve quality of ontologies, while considerably reducing the amount of work that has to be completed by human experts.

Although the automatic ontology inspection methods, such as the OBO Dashboard, do help with the problem of ontology quality, they cannot solve it on their own. There will always be limits to the scope of checks performed by automated, general tools. Thus, it is vital to pay special attention to the issues observed during reuse, and \emph{continuously} integrate resulting observations to improve the ontologies. Reusability is, arguably, the \emph{essence} of the Linked Open Data vision. Therefore, it should be the core driving factor for ontology development. It was shown that using multiple ontologies together is particularly hard, with numerous, still largely unresolved, issues regarding cross-ontology references. Thus, this topic deserves special attention in future research. 

Overall, it was demonstrated how reusing ontologies, especially in demanding applications, serves as the real test of their quality. The purpose of the performed experiments is not to present them as general approaches to ontology quality assurance. Rather, they exemplify the process of continuous ontology reuse, evaluation, and improvement.

\section*{Acknowledgments}

We would like to thank Anastasiya Danilenka and Jan Sawicki for their invaluable help in reviewing CSO's entities.

\bibliographystyle{vancouver}
\bibliography{bib/bibliography}

\begin{thebibliography}{10}

\bibitem{noy2001}
Noy N, Mcguinness D.
\newblock Ontology Development 101: A Guide to Creating Your First Ontology.
\newblock Knowledge Systems Laboratory. 2001 01;32.

\bibitem{bizer2011linked}
Bizer C, Heath T, Berners-Lee T.
\newblock Linked data: The story so far.
\newblock In: Semantic services, interoperability and web applications:
  emerging concepts. IGI global; 2011. p. 205-27.

\bibitem{fernandez2019ontologies}
Fern{\'a}ndez-L{\'o}pez M, Poveda-Villal{\'o}n M, Su{\'a}rez-Figueroa MC,
  G{\'o}mez-P{\'e}rez A.
\newblock Why are ontologies not reused across the same domain?
\newblock Journal of Web Semantics. 2019;57:100492.

\bibitem{cota2020landscape}
Cota G, et~al.
\newblock The landscape of ontology reuse approaches.
\newblock Appl Practices Ontol Des, Extraction, Reason. 2020;49:21.

\bibitem{brank2005survey}
Brank J, Grobelnik M, Mladenic D.
\newblock A survey of ontology evaluation techniques.
\newblock In: Proceedings of the conference on data mining and data warehouses
  (SiKDD 2005). Citeseer Ljubljana, Slovenia; 2005. p. 166-70.

\bibitem{raad2015survey}
Raad J, Cruz C.
\newblock A survey on ontology evaluation methods.
\newblock In: Proceedings of the International Conference on Knowledge
  Engineering and Ontology Development, part of the 7th International Joint
  Conference on Knowledge Discovery, Knowledge Engineering and Knowledge
  Management; 2015. p. 179-86.

\bibitem{amith2018assessing}
Amith M, He Z, Bian J, Lossio-Ventura JA, Tao C.
\newblock Assessing the practice of biomedical ontology evaluation: Gaps and
  opportunities.
\newblock Journal of biomedical informatics. 2018;80:1-13.

\bibitem{al2019towards}
Al-Sayed MM, Hassan HA, Omara FA.
\newblock Towards evaluation of cloud ontologies.
\newblock Journal of Parallel and Distributed Computing. 2019;126:82-106.

\bibitem{casellas_eval}
Casellas N.
\newblock Ontology Evaluation through Usability Measures.
\newblock In: Meersman R, Herrero P, Dillon T, editors. On the Move to
  Meaningful Internet Systems: OTM 2009 Workshops. Berlin, Heidelberg: Springer
  Berlin Heidelberg; 2009. p. 594-603.

\bibitem{laadhar2020investigating}
Laadhar A, Abrah{\~a}o E, Jonquet C.
\newblock Investigating one million {XR}efs in thirthy ontologies from the
  {OBO} world.
\newblock In: 11th International Conference on Biomedical Ontologies ({ICBO});
  2020. p. G.1-12.

\bibitem{szmeja2013reengineering}
Szmeja P, Wasielewska K, Ganzha M, Drozdowicz M, Paprzycki M, Fidanova S,
  et~al.
\newblock Reengineering and extending the {A}gents in {G}rid {O}ntology.
\newblock In: International Conference on Large-Scale Scientific Computing.
  Springer; 2013. p. 565-73.

\bibitem{ganzha2017vocabulary}
Ganzha M, Paprzycki M, Pawlowski W, Szmeja P, Wasielewska K.
\newblock Towards Common Vocabulary for IoT Ecosystems - preliminary
  Considerations.
\newblock In: Nguyen NT, Tojo S, Nguyen LM, Trawinski B, editors. Intelligent
  Information and Database Systems - 9th Asian Conference, {ACIIDS} 2017,
  Kanazawa, Japan, April 3-5, 2017, Proceedings, Part {I}. vol. 10191 of
  Lecture Notes in Computer Science; 2017. p. 35-45.
\newblock Available from: \url{https://doi.org/10.1007/978-3-319-54472-4\_4}.

\bibitem{wasielewska2021enterprise}
Wasielewska-Michniewska K, Ganzha M, Paprzycki M, Denisiuk A.
\newblock Application of ontologies in the enterprise -- overview and critical
  analysis.
\newblock In: Proceedings of the Third International Conference on Information
  Management and Machine Intelligence: ICIMMI 2021, Algorithms for Intelligent
  Systems (AIS), TO APPEAR. Springer; 2022. .

\bibitem{sowinski2022topical}
Sowi{\'n}ski P, Wasielewska-Michniewska K, Ganzha M, Paprzycki M.
\newblock Topical Classification of Food Safety Publications with a Knowledge
  Base.
\newblock arXiv preprint arXiv:220100374. 2022.

\bibitem{smith2007obo}
Smith B, Ashburner M, Rosse C, Bard J, Bug W, Ceusters W, et~al.
\newblock The {OBO Foundry}: coordinated evolution of ontologies to support
  biomedical data integration.
\newblock Nature biotechnology. 2007;25(11):1251-5.

\bibitem{donnelly2006snomed}
Donnelly K, et~al.
\newblock {SNOMED-CT}: {T}he advanced terminology and coding system for
  {eHealth}.
\newblock Studies in health technology and informatics. 2006;121:279.

\bibitem{jackson2019robot}
Jackson RC, Balhoff JP, Douglass E, Harris NL, Mungall CJ, Overton JA.
\newblock {ROBOT}: a tool for automating ontology workflows.
\newblock BMC bioinformatics. 2019;20(1):1-10.

\bibitem{Jackson2021OBOFI}
Jackson RC, Matentzoglu N, Overton JA, Vita R, Balhoff JP, Buttigieg PL, et~al.
\newblock {OBO Foundry} in 2021: operationalizing open data principles to
  evaluate ontologies.
\newblock Database: The Journal of Biological Databases and Curation.
  2021;2021.

\bibitem{moreira2009ncbo}
Moreira D, Mungall C, Shah N, Aitken S, Richter JD, Redmond T, et~al.
\newblock The {NCBO OBOF} to {OWL} Mapping.
\newblock Nature Precedings. 2009:1-1.

\bibitem{Miles:09:SSK}
Miles A, Bechhofer S.
\newblock {SKOS} Simple Knowledge Organization System Reference.
\newblock W3C; 2009.

\bibitem{charles_tapley_hoyt_2020_4404608}
Hoyt CT. An integrative registry of biological databases, ontologies, and
  nomenclatures; v0.0.6. Zenodo; 2020.
\newblock Available from: \url{https://doi.org/10.5281/zenodo.4404608}.

\bibitem{courtot2011mireot}
Courtot M, Gibson F, Lister AL, Malone J, Schober D, Brinkman RR, et~al.
\newblock {MIREOT}: The minimum information to reference an external ontology
  term.
\newblock Applied Ontology. 2011;6(1):23-33.

\bibitem{salatino2018computer}
Salatino AA, Thanapalasingam T, Mannocci A, Osborne F, Motta E.
\newblock The {Computer Science Ontology}: a large-scale taxonomy of research
  areas.
\newblock In: International Semantic Web Conference. Springer; 2018. p.
  187-205.

\bibitem{auer2007dbpedia}
Auer S, Bizer C, Kobilarov G, Lehmann J, Cyganiak R, Ives Z.
\newblock {DB}pedia: A nucleus for a web of open data.
\newblock In: The semantic web. Springer; 2007. p. 722-35.

\bibitem{vrandevcic2014wikidata}
Vrande{\v{c}}i{\'c} D, Kr{\"o}tzsch M.
\newblock Wikidata: a free collaborative knowledgebase.
\newblock Communications of the ACM. 2014;57(10):78-85.

\bibitem{tanon2020yago}
Tanon TP, Weikum G, Suchanek F.
\newblock {YAGO} 4: A reason-able knowledge base.
\newblock In: European Semantic Web Conference. Springer; 2020. p. 583-96.

\bibitem{osborne2015klink}
Osborne F, Motta E.
\newblock Klink-2: integrating multiple web sources to generate semantic topic
  networks.
\newblock In: International Semantic Web Conference. Springer; 2015. p. 408-24.

\bibitem{ashburner2000gene}
Ashburner M, Ball CA, Blake JA, Botstein D, Butler H, Cherry JM, et~al.
\newblock Gene {O}ntology: tool for the unification of biology.
\newblock Nature genetics. 2000;25(1):25-9.

\bibitem{jaradeh2019open}
Jaradeh MY, Oelen A, Farfar KE, Prinz M, D'Souza J, Kismih{\'o}k G, et~al.
\newblock {Open Research Knowledge Graph}: next generation infrastructure for
  semantic scholarly knowledge.
\newblock In: Proceedings of the 10th International Conference on Knowledge
  Capture; 2019. p. 243-6.

\bibitem{hagberg2008exploring}
Hagberg A, Swart P, S~Chult D.
\newblock Exploring network structure, dynamics, and function using {NetworkX}.
\newblock Los Alamos National Lab.(LANL), Los Alamos, NM (United States); 2008.

\bibitem{BollobasBela2013MGT}
Bollobas B, Gehring FW, Halmos PR.
\newblock Modern Graph Theory. vol. 184 of Graduate Texts in Mathematics.
\newblock New York, NY: Springer; 2013.

\bibitem{cordasco2010community}
Cordasco G, Gargano L.
\newblock Community detection via semi-synchronous label propagation
  algorithms.
\newblock In: 2010 IEEE international workshop on: business applications of
  social network analysis (BASNA). IEEE; 2010. p. 1-8.

\bibitem{bechhofer2004owl}
Bechhofer S, Van~Harmelen F, Hendler J, Horrocks I, McGuinness DL,
  Patel-Schneider PF, et~al.
\newblock {OWL} web ontology language reference.
\newblock W3C recommendation. 2004;10(2):1-53.

\bibitem{ghazvinian2009creating}
Ghazvinian A, Noy NF, Musen MA.
\newblock Creating mappings for ontologies in biomedicine: simple methods work.
\newblock In: AMIA Annual Symposium Proceedings. vol. 2009. American Medical
  Informatics Association; 2009. p. 198.

\bibitem{song2020mpnet}
Song K, Tan X, Qin T, Lu J, Liu TY.
\newblock {MPN}et: Masked and permuted pre-training for language understanding.
\newblock arXiv preprint arXiv:200409297. 2020.

\bibitem{han2020wikicssh}
Han K, Yang P, Mishra S, Diesner J.
\newblock WikiCSSH: extracting computer science subject headings from
  Wikipedia.
\newblock In: ADBIS, TPDL and EDA 2020 Common Workshops and Doctoral
  Consortium. Springer; 2020. p. 207-18.

\end{thebibliography}

\end{document}